\documentclass[copyright,creativecommons]{eptcs}

\usepackage{etex}

\usepackage{wrapfig}
\usepackage{times}
\usepackage{proof}
\usepackage{amssymb}
\usepackage{amsmath}
\usepackage{url}
\usepackage{color}
\usepackage{multicol}
\usepackage{multirow}
\usepackage{wasysym}
\usepackage{stmaryrd}
\usepackage{algorithmicx}
\usepackage[noend]{algpseudocode}
\usepackage{algorithm}
\usepackage{pgfplots}

\usepackage[inline]{enumitem}

\usepackage{siunitx}

\newcommand{\leaveout}[1]{}

\usepackage{adjustbox}
\usepackage{array}
\newcolumntype{R}[2]{%
    >{\adjustbox{angle=#1,lap=\width-(#2)}\bgroup}%
    l%
    <{\egroup}%
}

\title{Boldly Going Where No Prover Has Gone Before }
\author{
 Giles Reger
\institute{University of Manchester, Manchester, UK }
\email{giles.reger@manchester.ac.uk}
}
\def\titlerunning{Boldly Going Where No Prover Has Gone Before }
\def\authorrunning{Giles Reger}

\begin{document}
\maketitle
\begin{abstract}
I argue that the most interesting goal facing researchers in automated reasoning is being able to solve problems that cannot currently be solved by existing tools and methods. This may appear obvious, and is clearly not an original thought, but focusing on this as a primary goal allows us to examine other goals in a new light. 
Many successful theorem provers employ a portfolio of different methods for solving problems. This changes the landscape on which we perform our research: solving problems that can already be solved may not improve the state of the art and a method that can solve a handful of problems unsolvable by current methods, but generally performs poorly on most problems, can be very useful. 
We acknowledge that forcing new methods to compete against portfolio solvers can stifle innovation. However, this is only the case when comparisons are made at the level of total problems solved. We propose a movement towards focussing on unique solutions in evaluation and competitions 
i.e.\ measuring the potential contribution to a portfolio solver. 
This state of affairs is particularly prominent in first-order logic, which is undecidable. When reasoning in a decidable logic there can be a focus on optimising a decision procedure and measuring average solving times. But in a setting where solutions are difficult to find, average solving times lose meaning, and whilst improving the efficiency of a technique can move potential solutions within acceptable time limits, in general, complementary strategies may be more successful. 
\end{abstract}

\section{Setting the Scene}


In this discussion we consider the reasoning problem of taking a formula and answering whether the formula is satisfiable or not\footnote{In the case of first-order theorem proving this is usually phrased as checking validity via unsatisfiability of the negation.} . In some cases, this problem may be decidable (e.g.\ for propositional logic) and in more interesting cases (at least for this paper) it will not be (e.g.\ for first-order logic, or theories of arithmetic). This represents a large and diverse set of research areas and realistically our focus is on theorem proving in first-order logic (our main area of experience), however we attempt to argue more generally.

Solutions to the above problem take the form of \emph{proof search} where we apply the term liberally to describe, for example, both the approach of searching a finite set of possible solutions (as in SAT) or the approach of searching an infinite clause space (as in proving for first-order logic). Even in the decidable cases, this proof search is exploring a huge space that is rarely structured in such a way that a single search strategy will lead to a solution (within the given resources). This leads to the introduction of \emph{heuristics} to guide proof search in directions likely to contain a solution. Such heuristics may prefer different parts of the search space and could give up qualities of proof search usually considered important, such as \emph{completeness}, with the aim of finding a solution more quickly.

To handle the above issue, a common solution is the use of \emph{portfolio solvers} that combine different strategies for exploring the solution space by trying a sequence of different approaches (we do not conflate this approach with \emph{collaborative} approaches, which we discuss shortly). Note that this description does not differentiate between differing proof search strategies implemented within a single solver and the combination of distinct solvers implementing fundamentally different approaches. Practically there is clearly a difference, but in the limit one could envisage a solver implementing all ideas (in Vampire~\cite{DBLP:conf/cav/KovacsV13} we have two saturation-based proof search implementations based on resolution and InstGen~\cite{Korovin2013} and a finite model finder~\cite{DBLP:conf/sat/Reger0V16}).

The typical portfolio approach is to perform \emph{time slicing} e.g.\ to slice up the available solving time between different strategies (first introduced in Gandalf \cite{Tammet1997}). This approach is based on the hypothesis (which we revisit later) that {\bf if a solution exists then a quick solution exists}, which implies that it is better, for example, to try 100 \emph{complementary} strategies for one second rather than one strategy for 100 seconds. We highlight the term complementary as this approach relies on the identification of strategies that are likely to explore different parts of the search space. Finally, we note that this time slicing approach is straightforwardly parallelisable or at least made concurrent where different strategies are interleaved.

In addition to portfolio approaches, solvers can be combined to achieve a synergistic effect. Collaboration can have two forms. A problem can be split into smaller parts and different approaches collaborate by focussing on different parts, or each different approach considers the whole problem and there is collaboration at the level of things learned that are, in general, helpful for proof search.  An example of the first kind would be the Clause-Diffusion method \cite{bonacina1995clause} for parallel proof search.  Examples of the second kind include the historic \emph{Team-Work} method \cite{10.1007/3-540-61708-6_45} for clause sharing (among many others), or our more recent work \cite{DBLP:conf/cade/RegerTV15} utilising the AVATAR architecture in Vampire \cite{DBLP:conf/cade/RegerSV15,DBLP:conf/cav/Voronkov14} to share knowledge about the search space between concurrent proof attempts. The distinction is not always clear; for example, in the typical Nelson-Oppen SMT solver \cite{Nelson:1979:SCD:357073.357079} decision procedures are combined to solve different parts of the problem based on their signature, thus splitting the problem up but also communicating disjunctions of equalities over shared variables.

A key advantage of both portfolio and collaborative approaches is that strategies particularly suited for a specialised set of difficult problems can be combined to give a more general approach. 
This relies on techniques for deciding which strategies to apply based either on (syntactic or semantic) characteristics of a problem, or on some user guidance (e.g.\ if the user thinks they know the answer they can already restrict search to strategies optimised for satisfiability or unsatisfiability).

%
%
%
%

\section{The Challenge: Solving The Unsolved}

Our position is that the main and overriding goal of research on this (very general) problem is, or at least should be, to be able to {\bf solve as many problems that people care about as possible}. This statement has two parameters that need addressing. Firstly, how do we characterise those problems that people care about and, secondly, what does it mean to solve a problem e.g.\ under what resource constraints? We argue that important problems should be motivated by some application and what counts as a solution should be tied to this application. For example, many program analysis tools that pass reasoning tasks to solvers give very short time limits and solutions should be found within these limits to `count'.

The above goal appears self-evident but acknowledging that this should be an overriding goal (a matter of debate perhaps) might change how we approach research in the field. Indeed, it suggests the following activities are important:
\begin{itemize}
	\item Extending existing techniques to promote complementary proof search
	\item Identifying characterisable sets of problems that are currently difficult to solve
	\item Developing specialised solutions to such sets
	\item Exploring methods for identifying and combining diverse strategies
	\item Developing methods for collaboration between approaches
\end{itemize}
These are already common activities within the field. 
However, our position also suggests that some activities are less desirable. For example, it does little to help a portfolio approach to improve an existing method that is not already the best method for a particular set of problems, unless the improvements demonstrably make it such. In addition, improving average solving times is, in itself, not helpful to this goal as it tells us little about the contribution to portfolio solving. 
Although we acknowledge that there are important research activities complementary to the above stated goal (e.g.\ advancing deeper theoretical understanding) which require different forms of research.

\section{(Re)thinking about Evaluation}

We acknowledge that taking the above approach can have potential shortcomings that stifle innovation as comparing a new technique against a portfolio approach makes it unlikely that the new approach can be shown to be a direct improvement. However, we believe that this reflects on poor evaluation criteria, rather than the portfolio approach itself. Therefore, we suggest some evaluation activities (see also \cite{Vampire2014and2015:Challenges_of_Evaluating_New}) that could help promote working towards the above identified goal:
\begin{itemize}
	\item It is important that the benchmark problems used for evaluation reflect the problems that people care about, particularly the hard ones that are difficult to solve
	\item Evaluation should focus on the contribution of a new technique to the union of possible techniques. This means that:
	\begin{itemize}
		\item Experiments should focus on short solving times, based on the hypothesis (see below) that if a solution exists a quick solution exists and hence this quick solution should be preferred for a schedule of strategies
		\item Results should identify the (number of) unique problems solved by each approach (or subset of approaches)
		\item Results should consider how to build schedules of strategies from the experimental results to obtain the best (resource constrained) results overall 
		\item When comparing sets of proof search parameters it is useful to consider the contribution of a single parameter value to the `best' schedule of strategies
	\end{itemize}
	\item Where possible, the problems for which an approach is particularly suited, or not suited, should be characterised
	\item Ideally, competitions should focus on the sets of problems that are unsolved, creating challenges to the community to solve those problems, rather than focussing on how well solvers can tackle problems known to be solvable.
\end{itemize}


Finally, the natural conclusion of this argument might appear to be that we should have a single prover capable of handling everything and representing the union of the state of the art. This is not our intent. We believe that competition is still an important component in driving research forward. In particular, diverse approaches are those that are most likely to be complementary in the end. 


\section{Quick Solutions Exist (Most of the Time)}

The previous discussion refers to a hypothesis that if a solution exists a quick solution exists. We would caveat this with `most of the time' -- there are clearly difficult problems that are just difficult however you approach them but it is our experience that the majority of problems (notably theorems) can be solved quickly if they can be solved at all. 

We present some evidence for this hypothesis beyond our appeal to folklore but do not claim that this evidence is substantially rigorous. We have inspected a large data set of results kindly provided by Martin Suda\footnote{This can be provided upon request. It is not currently hosted online anywhere due to its size.} running our Vampire theorem prover (for 60 seconds) on all relevant problems in TPTP using 801 different strategies employed in competitions (CASC and SMT-COMP) over the last few years (this experiment consists of significantly more than 10 million runs of Vampire).

Out of 14,722 solved problems, 13,756 (96\%) have a solution taking less than 10 seconds, 12,390 (87\%) have a solution taking less than 1 second, and 9690 (68\%) have a solution taking less than 0.1 seconds. But perhaps these problems are generally easy. However, we also compared the maximum and mean solution times in the following table. Each cell can be read as \emph{When the fastest solution is at most x seconds the mean/max solution takes more than y seconds}. Let us take one cell as an example. There are 8106 problems were some strategy takes more than 30 seconds but the fastest solution takes less than a second. For those problems (more than half of those we solve) picking the `wrong' strategy can make a substantial difference to how long it takes to solve the problem. For around 10\% of the problems the mean strategy took over 10 seconds whilst the best strategy took less than a second i.e. on average we would pick the `wrong' strategy.

\begin{center}
\vspace{1em}
\begin{tabular}{ccc|ccc}
&&& \multicolumn{3}{c}{Faster solution in at most (seconds)} \\
&&& 0.1 & 1 & 10 \\ \hline
 \multirow{6}{*}{\rotatebox[origin=c]{90}{\parbox[c]{3cm}{\centering Mean/Max solution takes more than}}}
& \multirow{3}{*}{\rotatebox[origin=c]{90}{\parbox[c]{1cm}{\centering Mean}}}
 & 
 1 &4224~(29\%) & 6727~(47\%) & - \\
& &10 & 292~(0\%) & 1442~(10\%) & 2476~(17\%) \\
& &30 & 1~(0\%) & 7~(0\%) & 137~(0\%) 
 \\ \cline{2-6}
&  \multirow{3}{*}{\rotatebox[origin=c]{90}{\parbox[c]{1cm}{\centering Max}}}
 & 
 1 &  7421~(52\%) & 9992~(70\%) & - \\
& &10 &6499~(49\%) & 9036~(63\%) & 10228~(72\%)\\
& &30 &5682~(40\%) & 8106~(57\%) & 9158~(64\%) \\
\end{tabular}
\vspace{1em}
\end{center}

We suggest that this data supports our hypothesis that quick solutions exist and can act as evidence for prioritising the search for complementary approaches. Although we should not ignore those problems unsolved within 60 seconds or those only solved `slowly' (e.g. the 4\% or 13\% of problems taking more than 10 or 1 second respectively).

We have a corollary hypothesis for which we do not present data. That is, that if a solution cannot be found quickly it is unlikely to be found -- the longer we wait the less likely we are to get a result e.g. most proof search diverges.



%


\section{Concluding Remark}

In this short paper we have argued for the portfolio approach as it supports what we see as the main goal of automated reasoning research. This argument might seem oversimplified but we hope that it encourages lively discussion at ARCADE and potentially leads to changes in how we evaluate our research.

Finally, I note that this perspective was motivated by a different point of view communicated at the previous ARCADE workshop that communicates a different message. I encourage the reader to read this other work \cite{ARCADE2017:Do_Portfolio_Solvers_Harm} to gain a balanced view on the topic.


\bibliography{bib}

\begin{thebibliography}{10}
\providecommand{\bibitemdeclare}[2]{}
\providecommand{\surnamestart}{}
\providecommand{\surnameend}{}
\providecommand{\urlprefix}{Available at }
\providecommand{\url}[1]{\texttt{#1}}
\providecommand{\href}[2]{\texttt{#2}}
\providecommand{\urlalt}[2]{\href{#1}{#2}}
\providecommand{\doi}[1]{doi:\urlalt{http://dx.doi.org/#1}{#1}}
\providecommand{\bibinfo}[2]{#2}

\bibitemdeclare{article}{bonacina1995clause}
\bibitem{bonacina1995clause}
\bibinfo{author}{Maria~Paola \surnamestart Bonacina\surnameend} \&
  \bibinfo{author}{Jieh \surnamestart Hsiang\surnameend}
  (\bibinfo{year}{1995}): \emph{\bibinfo{title}{The Clause-Diffusion
  methodology for distributed deduction}}.
\newblock {\sl \bibinfo{journal}{Fundamenta Informaticae}}
  \bibinfo{volume}{24}(\bibinfo{number}{1, 2}), pp. \bibinfo{pages}{177--207},
  \doi{10.3233/FI-1995-24128}.

\bibitemdeclare{inproceedings}{10.1007/3-540-61708-6_45}
\bibitem{10.1007/3-540-61708-6_45}
\bibinfo{author}{J{\"o}rg \surnamestart Denzinger\surnameend} \&
  \bibinfo{author}{Martin \surnamestart Kronenburg\surnameend}
  (\bibinfo{year}{1996}): \emph{\bibinfo{title}{Planning for distributed
  theorem proving: The teamwork approach}}.
\newblock In \bibinfo{editor}{G{\"u}nther \surnamestart G{\"o}rz\surnameend} \&
  \bibinfo{editor}{Steffen \surnamestart H{\"o}lldobler\surnameend}, editors:
  {\sl \bibinfo{booktitle}{KI-96: Advances in Artificial Intelligence}},
  \bibinfo{publisher}{Springer Berlin Heidelberg}, \bibinfo{address}{Berlin,
  Heidelberg}, pp. \bibinfo{pages}{43--56},
  \doi{10.1007/3-540-61708-6\_45}.

\bibitemdeclare{inbook}{Korovin2013}
\bibitem{Korovin2013}
\bibinfo{author}{Konstantin \surnamestart Korovin\surnameend}
  (\bibinfo{year}{2013}): \emph{\bibinfo{title}{Inst-Gen -- A Modular Approach
  to Instantiation-Based Automated Reasoning}}, pp. \bibinfo{pages}{239--270}.
\newblock \bibinfo{publisher}{Springer Berlin Heidelberg},
  \bibinfo{address}{Berlin, Heidelberg}, \doi{10.1007/978-3-642-37651-1\_10}.

\bibitemdeclare{inproceedings}{DBLP:conf/cav/KovacsV13}
\bibitem{DBLP:conf/cav/KovacsV13}
\bibinfo{author}{Laura \surnamestart Kov{\'{a}}cs\surnameend} \&
  \bibinfo{author}{Andrei \surnamestart Voronkov\surnameend}
  (\bibinfo{year}{2013}): \emph{\bibinfo{title}{First-Order Theorem Proving and
  Vampire}}.
\newblock In: {\sl \bibinfo{booktitle}{Computer Aided Verification - 25th
  International Conference, {CAV} 2013, Saint Petersburg, Russia, July 13-19,
  2013. Proceedings}}, pp. \bibinfo{pages}{1--35},
  \doi{10.1007/978-3-642-39799-8\_1}.

\bibitemdeclare{article}{Nelson:1979:SCD:357073.357079}
\bibitem{Nelson:1979:SCD:357073.357079}
\bibinfo{author}{Greg \surnamestart Nelson\surnameend} \&
  \bibinfo{author}{Derek~C. \surnamestart Oppen\surnameend}
  (\bibinfo{year}{1979}): \emph{\bibinfo{title}{Simplification by Cooperating
  Decision Procedures}}.
\newblock {\sl \bibinfo{journal}{ACM Trans. Program. Lang. Syst.}}
  \bibinfo{volume}{1}(\bibinfo{number}{2}), pp. \bibinfo{pages}{245--257},
  \doi{10.1145/357073.357079}.

\bibitemdeclare{inproceedings}{DBLP:conf/cade/RegerSV15}
\bibitem{DBLP:conf/cade/RegerSV15}
\bibinfo{author}{Giles \surnamestart Reger\surnameend}, \bibinfo{author}{Martin
  \surnamestart Suda\surnameend} \& \bibinfo{author}{Andrei \surnamestart
  Voronkov\surnameend} (\bibinfo{year}{2015}): \emph{\bibinfo{title}{Playing
  with {AVATAR}}}.
\newblock In: {\sl \bibinfo{booktitle}{Automated Deduction - {CADE-25} - 25th
  International Conference on Automated Deduction, Berlin, Germany, August 1-7,
  2015, Proceedings}}, pp. \bibinfo{pages}{399--415},
  \doi{10.1007/978-3-319-21401-6\_28}.

\bibitemdeclare{inproceedings}{Vampire2014and2015:Challenges_of_Evaluating_New}
\bibitem{Vampire2014and2015:Challenges_of_Evaluating_New}
\bibinfo{author}{Giles \surnamestart Reger\surnameend}, \bibinfo{author}{Martin
  \surnamestart Suda\surnameend} \& \bibinfo{author}{Andrei \surnamestart
  Voronkov\surnameend} (\bibinfo{year}{2016}): \emph{\bibinfo{title}{The
  Challenges of Evaluating a New Feature in Vampire}}.
\newblock In \bibinfo{editor}{Laura \surnamestart
  Kov\textbackslash{}'acs\surnameend} \& \bibinfo{editor}{Andrei \surnamestart
  Voronkov\surnameend}, editors: {\sl \bibinfo{booktitle}{Proceedings of the
  1st and 2nd Vampire Workshops}}, {\sl \bibinfo{series}{EPiC Series in
  Computing}}~\bibinfo{volume}{38}, \bibinfo{publisher}{EasyChair}, pp.
  \bibinfo{pages}{70--74}, \doi{10.29007/1ffk}.

\bibitemdeclare{inproceedings}{DBLP:conf/sat/Reger0V16}
\bibitem{DBLP:conf/sat/Reger0V16}
\bibinfo{author}{Giles \surnamestart Reger\surnameend}, \bibinfo{author}{Martin
  \surnamestart Suda\surnameend} \& \bibinfo{author}{Andrei \surnamestart
  Voronkov\surnameend} (\bibinfo{year}{2016}): \emph{\bibinfo{title}{Finding
  Finite Models in Multi-sorted First-Order Logic}}.
\newblock In: {\sl \bibinfo{booktitle}{Theory and Applications of
  Satisfiability Testing - {SAT} 2016 - 19th International Conference,
  Bordeaux, France, July 5-8, 2016, Proceedings}}, pp.
  \bibinfo{pages}{323--341}, \doi{10.1007/978-3-319-40970-2\_20}.

\bibitemdeclare{inproceedings}{DBLP:conf/cade/RegerTV15}
\bibitem{DBLP:conf/cade/RegerTV15}
\bibinfo{author}{Giles \surnamestart Reger\surnameend}, \bibinfo{author}{Dmitry
  \surnamestart Tishkovsky\surnameend} \& \bibinfo{author}{Andrei \surnamestart
  Voronkov\surnameend} (\bibinfo{year}{2015}):
  \emph{\bibinfo{title}{Cooperating Proof Attempts}}.
\newblock In: {\sl \bibinfo{booktitle}{Automated Deduction - {CADE-25} - 25th
  International Conference on Automated Deduction, Berlin, Germany, August 1-7,
  2015, Proceedings}}, pp. \bibinfo{pages}{339--355},
  \doi{10.1007/978-3-319-21401-6\_23}.

\bibitemdeclare{article}{Tammet1997}
\bibitem{Tammet1997}
\bibinfo{author}{Tanel \surnamestart Tammet\surnameend} (\bibinfo{year}{1997}):
  \emph{\bibinfo{title}{Gandalf}}.
\newblock {\sl \bibinfo{journal}{Journal of Automated Reasoning}}
  \bibinfo{volume}{18}(\bibinfo{number}{2}), pp. \bibinfo{pages}{199--204},
  \doi{10.1023/A:1005887414560}.

\bibitemdeclare{inproceedings}{DBLP:conf/cav/Voronkov14}
\bibitem{DBLP:conf/cav/Voronkov14}
\bibinfo{author}{Andrei \surnamestart Voronkov\surnameend}
  (\bibinfo{year}{2014}): \emph{\bibinfo{title}{{AVATAR:} The Architecture for
  First-Order Theorem Provers}}.
\newblock In: {\sl \bibinfo{booktitle}{Computer Aided Verification - 26th
  International Conference, {CAV} 2014, Held as Part of the Vienna Summer of
  Logic, {VSL} 2014, Vienna, Austria, July 18-22, 2014. Proceedings}}, pp.
  \bibinfo{pages}{696--710}, \doi{10.1007/978-3-319-08867-9\_46}.

\bibitemdeclare{inproceedings}{ARCADE2017:Do_Portfolio_Solvers_Harm}
\bibitem{ARCADE2017:Do_Portfolio_Solvers_Harm}
\bibinfo{author}{Christoph \surnamestart Weidenbach\surnameend}
  (\bibinfo{year}{2017}): \emph{\bibinfo{title}{Do Portfolio Solvers Harm?}}
\newblock In \bibinfo{editor}{Giles \surnamestart Reger\surnameend} \&
  \bibinfo{editor}{Dmitriy \surnamestart Traytel\surnameend}, editors: {\sl
  \bibinfo{booktitle}{ARCADE 2017. 1st International Workshop on Automated
  Reasoning: Challenges, Applications, Directions, Exemplary Achievements}},
  {\sl \bibinfo{series}{EPiC Series in Computing}}~\bibinfo{volume}{51},
  \bibinfo{publisher}{EasyChair}, pp. \bibinfo{pages}{76--81},
  \doi{10.29007/vpxm}.

\end{thebibliography}
\bibliographystyle{eptcs}

\end{document}